\documentclass[twocolumn,showpacs,preprintnumbers,amsmath,amssymb]{revtex4}

\usepackage{graphicx}
\usepackage{dcolumn}
\usepackage{bm}

\begin{document}


\title{Dynamics of Particles in Non Scaling FFAG Accelerators}

\author{James K. Jones}
  \email{james.jones@stfc.ac.uk}
\author{Bruno D. Muratori}%
 \email{bruno.muratori@stfc.ac.uk}
\author{Susan L. Smith}%
 \email{susan.smith@stfc.ac.uk}
\author{Stephan I. Tzenov}%
 \email{stephan.tzenov@stfc.ac.uk}

\affiliation{%
STFC Daresbury Laboratory, Daresbury, Warrington, Cheshire, WA4 4AD, United Kingdom
}%

\date{\today}

\begin{abstract}
Non scaling Fixed-Field Alternating Gradient (FFAG) accelerators have an unprecedented potential for muon acceleration, as well as for medical purposes based on carbon and proton hadron therapy. They also represent a possible active element for an Accelerator Driven Subcritical Reactor (ADSR). Starting from first principle the Hamiltonian formalism for the description of the dynamics of particles in non scaling FFAG machines has been developed. The stationary reference (closed) orbit has been found within the Hamiltonian framework. The dependence of the path length on the energy deviation has been described in terms of higher order dispersion functions. The latter have been used subsequently to specify the longitudinal part of the Hamiltonian. It has been shown that higher order phase slip coefficients should be taken into account to adequately describe the acceleration in non scaling FFAG accelerators. A complete theory of the fast (serpentine) acceleration in non scaling FFAGs has been developed. An example of the theory is presented for the parameters of the Electron Machine with Many Applications (EMMA), a prototype electron non scaling FFAG to be hosted at Daresbury Laboratory.
\end{abstract}

\maketitle

\section{\label{sec:level1}Introduction}

Fixed-Field Alternating Gradient (FFAG) accelerators were proposed half century ago \cite{KL,Kol,Sy,Ke}, when acceleration of electrons was first
demonstrated. These machines, which were intensively studied in the 1950s and 1960s but never progressed beyond the model stage, have in recent years become the focus of renewed attention. Acceleration of protons has been recently achieved at the KEK Proof-of-Principle (PoP) proton FFAG \cite{Ai}.

To avoid the slow crossing of betatron resonances associated with a typical low energy-gain per turn, the first FFAGs designed and constructed so far have been based on the "scaling" principle. The latter implies that the orbit shape and betatron tunes must be kept fixed during the acceleration process. Thus, magnets must be built with constant field index, while in the case of spiral-sector designs the spiral angle must be constant as well. Machines of this type use conventional magnets with the bending and focusing field being kept constant during acceleration. The latter alternate in sign, providing a more compact radial extension and consequently smaller aperture as compared to the AVF cyclotrons. The ring essentially consists of a sequence of short cells with very large periodicity.

Non scaling FFAG machines have until recently been considered as an alternative. The bending and the focusing is provided simultaneously by focusing and defocusing quadrupole magnets repeating in an alternating sequence. There is a number of advantages of the non scaling FFAG lattice as compared to the scaling one, among which are the relatively small transverse magnet aperture (tending to be much smaller than the one for scaling machines) and the lower field strength. Unfortunately this lattice leads to a large betatron tune variation across the required energy range for acceleration as opposed to the scaling lattice. As a consequence several resonances are crossed during the acceleration cycle, some of them nonlinear created by the magnetic field imperfections, as well as half-integer and integer ones. A possible bypass to this problem is the rapid acceleration (of utmost importance for muons), which allows betatron resonances no time to essentially damage beam quality.

Because non scaling FFAG accelerators have otherwise very desirable features, it is important to investigate analytically and numerically some of the peculiarities of the beam dynamics, the new type of fast acceleration regime (so-called serpentine acceleration) and the effects of crossing of linear as well as nonlinear resonances. Moreover, it is important to examine the most favorable phase at which the cavities need to be set for the optimal acceleration. Some of these problems will be discussed in the present paper.

An example of the theory developed here is presented for the parameters of the Electron Machine with Many Applications (EMMA) \cite{emma}, a prototype electron non scaling FFAG to be hosted at Daresbury Laboratory. The Accelerators and Lasers In Combined Experiments (ALICE) accelerator \cite{alice} is used as an injector to the EMMA ring. The energy delivered by this injector can vary from a $10$ to $20$ MeV single bunch train with a bunch charge of $16$ to $32$ pC at a rate of $1$ to $20$ Hz. ALICE is presently designed to deliver bunches which are around $4$ ps and $8.35$ MeV from the exit of the booster of its injector line. These are then accelerated to $10$ or $20$ MeV in the main ALICE linac after which they are sent to the EMMA injection line. The EMMA injection line ends with a septum for injection into the EMMA ring itself followed by two kickers so as to direct the beam onto the correct, energy dependent, trajectory. After circulation in the EMMA ring, the electron bunches are extracted using what is almost a mirror image of the injection setup with two kickers followed by an extraction septum. The beam is then transported to a diagnostic line whose purpose it is to analyze in as much detail as possible the effect the non scaling FFAG has had on the bunch.

The paper is organized as follows. Firstly, we review some generalities and first principles of the Hamiltonian formalism \cite{Tzenov} suitably modified to cover the case of a non scaling FFAG lattice. Subsequently the synchrobetatron framework is applied to determine the energy dependent reference orbit. Stability of motion about the stationary reference orbit is described in terms of betatron oscillations with energy dependent Twiss parameters and betatron tunes. Dispersion, measuring the effect of energy variation on the path length along the reference orbit is an essential feature of non scaling FFAGs. Within the developed synchrobetatron formalism higher order dispersion functions have been introduced and their contribution to the longitudinal dynamics has been further analyzed. Finally, a complete description of the so-called serpentine acceleration in non scaling lepton FFAGs is given together with conclusions. The calculations of the reference orbit and phase stability are detailed in the appendices.

\section{Generalities and First Principles}

Let the ideal (design) trajectory of a particle in an accelerator be a planar curve with curvature $K$. The Hamiltonian describing the motion of a particle in a natural coordinate system attached to the orbit thus defined is \cite{Tzenov}:
\begin{widetext}
\begin{equation}
H = - {\left( 1 + Kx \right)} {\sqrt{{\frac {{\left( {\mathcal{H}} - q \varphi \right)}^2} {c^2}} - m_{p0}^2 c^2 - {\left( P_x - q A_x \right)}^2 - {\left( P_z - q A_z \right)}^2}} - q {\left( 1 + Kx \right)} A_s, \label{BHamilt}
\end{equation}
\end{widetext}
\noindent where $m_{p0}$ is the rest mass of the particle. The guiding magnetic field can be represented as a gradient of a function $\psi {\left( x, z; s \right)}$
\begin{equation}
\mathbf{B} = \nabla \psi, \label{Mfield}
\end{equation}
\noindent where the latter satisfies the Laplace equation
\begin{equation}
\nabla^2 \psi = 0. \label{Poisson}
\end{equation}
\noindent Using the median symmetry of the machine, it is straightforward to show that $\psi$ can be written in the form
\begin{equation}
\psi = {\left( a_0 + a_1 x + {\frac {a_2 x^2} {2!}} + \dots \right)} z \nonumber
\end{equation}
\begin{equation}
- {\left( b_0 + b_1 x + {\frac {b_2 x^2} {2!}} + \dots \right)} {\frac {z^3} {3!}} + {\left( c_0 + c_1 x + \dots \right)} {\frac {z^5} {5!}} + \dots. \label{Psifunction}
\end{equation}
\noindent Inserting the above expression into the Laplace equation (\ref{Poisson}), one readily finds relations between the coefficients $b_k$ and $c_k$ on one hand and $a_k$ on the other:
\begin{equation}
b_0 = a_0^{\prime \prime} + K a_1 + a_2, \label{Coefficb0}
\end{equation}
\begin{equation}
b_1 = - 2K a_0^{\prime \prime} - K^{\prime} a_0^{\prime} + a_1^{\prime \prime} - K^2 a_1 + K a_2 + a_3, \label{Coefficb1}
\end{equation}
\begin{equation}
b_2 = 6 K^2 a_0^{\prime \prime} + 6K K^{\prime} a_0^{\prime} - 4K a_1^{\prime \prime} - 2 K^{\prime} a_1^{\prime} \nonumber
\end{equation}
\begin{equation}
+ a_2^{\prime \prime} + 2 K^3 a_1 - 2 K^2 a_2 + K a_3 + a_4, \label{Coefficb2}
\end{equation}
\begin{equation}
c_0 = b_0^{\prime \prime} + K b_1 + b_2. \label{Coefficc0}
\end{equation}
\noindent Prime in the above expressions implies differentiation with respect to the longitudinal coordinate $s$. The coefficients $a_k$ have a very simple meaning:
\begin{equation}
a_0 = {\left( B_z \right)}_{x,z=0}, \qquad a_1 = {\left( {\frac {\partial B_z} {\partial x}} \right)}_{x,z=0}, \nonumber
\end{equation}
\begin{equation}
a_2 = {\left( {\frac {\partial^2 B_z} {\partial x^2}} \right)}_{x,z=0}. \label{Coeffica}
\end{equation}
\noindent In other words, this implies that, provided the vertical component $B_z$ of the magnetic field and its derivatives with respect to the horizontal coordinate $x$ are known in the median plane, one can in principle reconstruct the entire field chart.

The vector potential $\mathbf{A}$ can be represented as
\begin{equation}
A_x = -z \overline{F} {\left( x, z; s \right)}, \quad A_z = x \overline{F} {\left( x, z; s \right)}, \quad A_s = \overline{G} {\left( x, z; s \right)}, \label{Vectorpot}
\end{equation}
\noindent where the Poincar${\grave{\rm e}}$ gauge condition
\begin{equation}
x A_x + z A_z = 0, \label{Gauge}
\end{equation}
\noindent written in the natural coordinate system has been used. From Maxwell's equation
\begin{equation}
\mathbf{B} = \nabla \times \mathbf{A}, \label{Maxwell}
\end{equation}
\noindent we obtain
\begin{equation}
2 \overline{F} + {\left( x \partial_x + z \partial_z \right)}
\overline{F} = B_s, \label{Eulerhomog1}
\end{equation}
\begin{equation}
{\frac {Kx} {1 + Kx}} \overline{G} + {\left( x \partial_x + z \partial_z \right)} \overline{G} = z B_x - x B_z. \label{Eulerhomog2}
\end{equation}
\noindent Applying Euler's theorem for homogeneous functions, we can write
\begin{equation}
\overline{F} = {\frac {1} {2}} B_s^{(0)} + {\frac {1} {3}} B_s^{(1)} + {\frac {1} {4}} B_s^{(2)} + \dots, \label{HomogeneousF}
\end{equation}
\begin{equation}
{\overline{G}}_u = {\left( 1 + {\frac {Kx} {2}} \right)} B_u^{(0)} + {\left( {\frac {1} {2}} + {\frac {Kx} {3}} \right)} B_u^{(1)} \nonumber
\end{equation}
\begin{equation}
+ {\left( {\frac {1} {3}} + {\frac {K x} {4}} \right)} B_u^{(2)} + \dots, \label{HomogeneousGu}
\end{equation}
\begin{equation}
{\overline{G}} = {\frac {z {\overline{G}}_x - x {\overline{G}}_z} {1 + Kx}}. \label{HomogeneousG}
\end{equation}
\noindent Here $u = (x, z)$ and $B_{\alpha}^{(k)}$ denotes homogeneous polynomials in $x$ and $z$ of order $k$, representing the corresponding parts of the components of the magnetic field $\mathbf{B} = {\left( B_x, B_z, B_s \right)}$. Thus, having found the magnetic field represented by equation (\ref{Psifunction}), it is straightforward to calculate the vector potential $\mathbf{A}$.

The accelerating field in AVF cyclotrons and FFAG machines can be represented by a scalar potential $\varphi$ (the
corresponding vector potential $\mathbf{A} = 0$). Due to the median symmetry, we have
\begin{equation}
\varphi = A_0 + A_1 x + {\frac {A_2 x^2} {2!}} + \dots \nonumber
\end{equation}
\begin{equation}
- {\left( B_0 + B_1 x + {\frac {B_2 x^2} {2!}} + \dots \right)} {\frac {z^2} {2!}} \nonumber
\end{equation}
\begin{equation}
+ {\left( C_0 + C_1 x + \dots \right)} {\frac {z^4} {4!}} + \dots. \label{Phifunction}
\end{equation}
\noindent Inserting the above expansion into the Laplace equation for $\varphi$, we obtain similar relations between $B_k$ and $C_k$ on one hand and $A_k$ on the other, which are analogous to those relating $b_k$, $c_k$ and $a_k$.

We consider the canonical transformation, specified by the generating function
\begin{equation}
S_2 {\left( x, z, {\cal T}, {\widehat{P}}_x, {\widehat{P}}_z, E; s \right)} = x {\widehat{P}}_x + z {\widehat{P}}_z + {\cal T} E \nonumber
\end{equation}
\begin{equation}
+ q \int {\rm d} {\cal T} \varphi {\left( x, z, {\cal T}; s \right)}, \label{Genfunction}
\end{equation}
\noindent where
\begin{equation}
{\cal T} = -t, \label{Sigma}
\end{equation}
\noindent is a canonical variable canonically conjugate to $\mathcal{H}$. The relations between the new and the old variables are
\begin{equation}
\widehat{u} = {\frac{\partial S_2}{\partial {\widehat{P}}_u}} = u, \qquad u = {\left( x, z \right)}, \qquad \widehat{\cal T} = {\frac{\partial S_2}{\partial E}} = {\cal T}, \label{Newoldcoord}
\end{equation}
\begin{equation}
P_u = {\frac{\partial S_2}{\partial u}} = {\widehat{P}}_u - q \int {\rm d} {\cal T} E_u {\left( x, z, {\cal T}; s \right)} \nonumber
\end{equation}
\begin{equation}
= {\widehat{P}}_u - q {\widetilde{E}}_u {\left( x, z, {\cal T}; s \right)}, \qquad E_u = - {\frac {\partial \varphi} {\partial u}}, \label{Newoldmom}
\end{equation}
\begin{equation}
\mathcal{H}= {\frac{\partial S_2}{\partial {\cal T}}} = E + q \varphi {\left( x, z, {\cal T}; s \right)} = m_{p0} \gamma c^2 + q \varphi {\left( x, z, {\cal T}; s \right)}. \label{Newoldener}
\end{equation}
\noindent The new Hamiltonian acquires now the form
\begin{widetext}
\begin{equation}
\widehat{H}= - {\left( 1 + K x \right)} {\sqrt{{\frac {E^2} {c^2}} - m_{p0}^2 c^2 - {\left( \widehat{P}_x - q \widetilde{E}_x - q A_x \right)}^2 - {\left( \widehat{P}_z - q \widetilde{E}_z - q A_z \right)}^2}} - q {\left( 1 + Kx \right)} {\left( A_s + \widetilde{E}_s \right)}, \label{BHamilto}
\end{equation}
\end{widetext}
\noindent where
\begin{equation}
\widetilde{E}_s = \int {\rm d} {\cal T} E_s {\left( x, z, {\cal T}; s \right)} \nonumber
\end{equation}
\begin{equation}
= - {\frac {1} {1 + Kx}} \int {\rm d} {\cal T} {\frac {\partial \varphi {\left( x, z, {\cal T}; s \right)}} {\partial s}}. \label{Efieldtilde}
\end{equation}
\noindent We introduce the new scaled variables
\begin{equation}
\widetilde{P}_u = {\frac {\widehat{P}_u} {p_0}} = {\frac {\widehat{P}_u} {m_{p0} c}}, \quad \Theta = c {\cal T}, \quad \gamma = {\frac {E} {E_p}} = {\frac {E} {m_{p0} c^2}}. \label{Scaledvar}
\end{equation}
\noindent The new scaled Hamiltonian can be expressed as
\begin{widetext}
\begin{equation}
\widetilde{H} = {\frac {\widehat{H}} {p_0}} = - {\left( 1 + K x \right)} {\sqrt{\gamma^2 - 1 - {\left( \widetilde{P}_x - \widetilde{q} \widetilde{E}_x - \widetilde{q} A_x \right)}^2 - {\left( \widetilde{P}_z - \widetilde{q} \widetilde{E}_z - \widetilde{q} A_z \right)}^2}} - \widetilde{q} {\left( 1 + K x \right)} {\left( A_s + \widetilde{E}_s \right)}, \label{SHamilt}
\end{equation}
\end{widetext}
\noindent where
\begin{equation}
\widetilde{q} = {\frac {q} {p_0}}. \label{Chargetilde}
\end{equation}
\noindent The quantities $\widetilde{E}_x$ and $\widetilde{E}_z$ can be neglected as compared to the components of the vector potential $\mathbf{A}$, so that
\begin{widetext}
\begin{equation}
\widetilde{H} = \beta \gamma {\left( 1 + K x \right)} {\left[ - {\sqrt{1 - {\left( \overline{P}_x - \overline{q} A_x \right)}^2 - {\left( \overline{P}_z - \overline{q} A_z \right)}^2}} - \overline{q} A_s \right]} - \widetilde{q} {\left( 1 + K x \right)} \widetilde{E}_s, \label{SHamilto}
\end{equation}
\end{widetext}
\noindent where now
\begin{equation}
\overline{q} = {\frac {q} {p}} = {\frac {q} {\beta \gamma p_0}}, \qquad \overline{P}_u = {\frac {\widehat{P}_u} {p}} = {\frac {\widehat{P}_u} {\beta \gamma p_0}}, \qquad u = (x, z). \label{Chargeoline}
\end{equation}
\noindent Since $\overline{P}_u$ and $u$ are small deviations, we can expand the square root in power series in the canonical variables $x$, $\overline{P}_x$ and $z$, $\overline{P}_z$. Tedious algebra yields
\begin{equation}
\widetilde{H} = \widetilde{H}_0 + \widetilde{H}_1 + \widetilde{H}_2 + \widetilde{H}_3 + \widetilde{H}_4 + \dots, \label{SHamiltdec}
\end{equation}
\begin{equation}
\widetilde{H}_0 = - \beta \gamma - \widetilde{q} {\left( 1 + Kx \right)} \widetilde{E}_s, \label{SHamiltdec0}
\end{equation}
\begin{equation}
\widetilde{H}_1 = \beta \gamma {\left( \overline{q} a_0 - K \right)} x, \label{SHamiltdec1}
\end{equation}
\begin{equation}
\widetilde{H}_2 = {\frac {\beta \gamma} {2}} {\left( {\overline{P}}_x^2 + {\overline{P}}_z^2 \right)} + {\frac {\widetilde{q}} {2}} {\left[ {\left( Ka_0 + a_1 \right)} x^2 - a_1 z^2 \right]}, \label{SHamiltdec2}
\end{equation}
\begin{widetext}
\begin{equation}
\widetilde{H}_3 = {\frac {\beta \gamma} {2}} Kx {\left( {\overline{P}}_x^2 + {\overline{P}}_z^2 \right)} + {\frac {\widetilde{q} a_0^{\prime} z} {3}} {\left( z \overline{P}_x - x \overline{P}_z \right)} + {\frac {\widetilde{q}} {3}} {\left[ {\left( K a_1 + {\frac {a_2} {2}} \right)} x^3 - {\left( K a_1 + a_2 + {\frac {b_0} {2}} \right)} x z^2 \right]}, \label{SHamiltdec3}
\end{equation}
\begin{equation}
\widetilde{H}_4 = {\frac {\beta \gamma} {8}} {\left( {\overline{P}}_x^2 + {\overline{P}}_z^2 \right)}^2 + {\frac {\widetilde{q} x z} {12}} {\left( K a_0^{\prime} + 3 a_1^{\prime} \right)} {\left( z \overline{P}_x - x \overline{P}_z \right)} + {\frac {\overline{q}^2 \beta \gamma a_0^{\prime 2} z^2} {18}} {\left( x^2 + z^2 \right)} \nonumber
\end{equation}
\begin{equation}
+ {\frac {\widetilde{q}} {4}} {\left[ {\left( {\frac {K a_2} {2}} + {\frac {a_3} {6}} \right)} x^4 - {\left( K a_2 + {\frac {a_3} {3}} + {\frac {K b_0} {2}} + {\frac {b_1} {2}} \right)} x^2 z^2 + {\frac {b_1} {6}} z^4 \right]}. \label{SHamiltdec4}
\end{equation}
\end{widetext}

The Hamiltonian decomposition (\ref{SHamiltdec}) represents the milestone of the synchrobetatron formalism. For instance, ${\widetilde{H}_0}$ governs the longitudinal motion, ${\widetilde{H}_1}$ describes linear coupling between longitudinal and transverse degrees of freedom and is the basic source of dispersion. The part ${\widetilde{H}_2}$ is responsible for linear betatron motion and chromaticity, while the remainder describes higher order contributions.

\section{The Synchro-Betatron Formalism and the Reference Orbit}

In the present paper we consider a FFAG lattice with polygonal structure. To define and subsequently calculate the stationary reference orbit, it is convenient to use a global Cartesian coordinate system whose origin is located in the center of the polygon. To describe step by step the fraction of the reference orbit related to a particular side of the polygon, we rotate each time the axes of the coordinate system by the polygon angle $\Theta_p = 2 \pi / N_L$, where $N_L$ is the number of sides of the polygon.

Let $X_e$ and $P_e$ denote the reference orbit and the reference momentum, respectively. The vertical component of the magnetic field in the median plane of a perfectly linear machine can be written as
\begin{equation}
B_z {\left( X_e; s \right)} = a_1 (s) {\left[ X_e - X_c - d (s) \right]}, \nonumber
\end{equation}
\begin{equation}
a_0 {\left( X_e; s \right)} = B_z {\left( X_e; s \right)}, \label{MagfldEMMA}
\end{equation}
\noindent where $s$ is the distance along the polygon side, and $X_c$ is the distance of the side of the polygon from the center of the machine
\begin{equation}
X_c = {\frac {L_p} {2 \tan (\Theta_p / 2)}}. \label{Polygside}
\end{equation}
\noindent Here $L_p$ is the length of the polygon side which actually represents the periodicity parameter of the lattice. Usually $X_c$ is related to an arbitrary energy in the range from injection to extraction energy. In the case of EMMA it is related to the 15 MeV orbit. The quantity $d (s)$ in equation (\ref{MagfldEMMA}) is the relative offset of the magnetic center in the quadrupoles with respect to the corresponding side of the polygon. In what follows [see equations (\ref{FocusQuadx}) and (\ref{DefocusQuadx})] $d_F$ corresponds to the offset in the focusing quadrupoles and $d_D$ corresponds to the one in the defocusing quadrupoles. Similarly, $a_F$ and $a_D$ stand for the particular value of $a_1$ in the focusing and the defocusing quadrupoles, respectively.

A design (reference) orbit corresponding to a local curvature $K {\left( X_e; s \right)}$ can be defined according to the relation
\begin{equation}
K {\left( X_e; s \right)} = {\frac {q} {p_0 \beta_e \gamma_e}} B_z {\left( X_e; s \right)}, \label{Referendes}
\end{equation}
\noindent where $\gamma_e$ is the energy of the reference particle. In terms of the reference orbit position $X_e (s)$ the equation for the curvature can be written as
\begin{equation}
X_e^{\prime \prime} = {\frac {q} {p_0 \beta_e \gamma_e}} {\left( 1 + X_e^{\prime2} \right)}^{3/2} B_z {\left( X_e; s \right)}, \label{Equarefrad}
\end{equation}
\noindent where the prime implies differentiation with respect to $s$.

To proceed further, we notice that equation (\ref{Equarefrad}) parameterizing the local curvature can be derived from an equivalent Hamiltonian
\begin{equation}
H_e {\left( X_e, P_e; s \right)} = - {\sqrt{\beta_e^2 \gamma_e^2 - P_e^2}} - {\widetilde{q}} \int {\rm d} X_e B_z {\left( X_e; s \right)}. \label{EquivHam}
\end{equation}
\noindent Taking into account Hamilton's equations of motion
\begin{equation}
X_e^{\prime} = {\frac {P_e} {\sqrt{\beta_e^2 \gamma_e^2 - P_e^2}}}, \qquad \qquad P_e^{\prime} = {\widetilde{q}} B_z {\left( X_e; s \right)}, \label{EquivHamEqMot}
\end{equation}
\noindent and using the relation
\begin{equation}
P_e = {\frac {\beta_e \gamma_e X_e^{\prime}} {\sqrt{1 + X_e^{\prime 2}}}}, \label{EquivRel}
\end{equation}
\noindent we readily obtain equation (\ref{Equarefrad}). Note also that the Hamiltonian (\ref{EquivHam}) follows directly from the scaled Hamiltonian (\ref{SHamilt}) with $x = 0$, ${\widetilde{P}}_x = P_e$, ${\widetilde{P}}_z = 0$, $A_x = A_z = 0$ and the accelerating cavities being switched off respectively.

Hamilton's equations of motion (\ref{EquivHamEqMot}) can be linearized and subsequently solved approximately by assuming that
\begin{equation}
P_e \ll \beta_e \gamma_e. \label{Approx}
\end{equation}
\noindent Thus, assuming electrons ($q = - e$), we have
\begin{equation}
P_e = \beta_e \gamma_e X_e^{\prime}, \qquad X_e^{\prime \prime} = - {\frac {e a_1 (s)} {p_0 \beta_e \gamma_e}} {\left( X_e - X_c - d (s) \right)}. \label{HamEquatLin}
\end{equation}
\noindent The three types of solutions to equations (\ref{HamEquatLin}) are as follows:

{\it Drift Space}
\begin{equation}
X_e = X_0 + {\frac {P_0} {\beta_e \gamma_e}} {\left( s - s_0 \right)}, \qquad \qquad P_e = P_0, \label{DriftSpace}
\end{equation}
\noindent where $X_0$ and $P_0$ are the initial position and reference momentum and $s$ is the distance in longitudinal direction.

{\it Focusing Quadrupole}
\begin{equation}
X_e = X_c + d_F + {\left( X_0 - X_c - d_F \right)} \cos \omega_F {\left( s - s_0 \right)} \nonumber
\end{equation}
\begin{equation}
+ {\frac {P_0} {\beta_e \gamma_e \omega_F}} \sin \omega_F {\left( s - s_0 \right)}, \label{FocusQuadx}
\end{equation}
\begin{equation}
P_e = - \beta_e \gamma_e \omega_F {\left( X_0 - X_c - d_F \right)} \sin \omega_F {\left( s - s_0 \right)} \nonumber
\end{equation}
\begin{equation}
+ P_0 \cos \omega_F {\left( s - s_0 \right)}, \label{FocusQuadp}
\end{equation}
\noindent where
\begin{equation}
\omega_F^2 = {\frac {e a_F} {p_0 \beta_e \gamma_e}}. \label{FocusFreq}
\end{equation}

{\it Defocusing Quadrupole}
\begin{equation}
X_e = X_c + d_D + {\left( X_0 - X_c - d_D \right)} \cosh \omega_D {\left( s - s_0 \right)} \nonumber
\end{equation}
\begin{equation}
+ {\frac {P_0} {\beta_e \gamma_e \omega_D}} \sinh \omega_D {\left( s - s_0 \right)}, \label{DefocusQuadx}
\end{equation}
\begin{equation}
P_e = \beta_e \gamma_e \omega_D {\left( X_0 - X_c - d_D \right)} \sinh \omega_D {\left( s - s_0 \right)} \nonumber
\end{equation}
\begin{equation}
+ P_0 \cosh \omega_D {\left( s - s_0 \right)}, \label{DefocusQuadp}
\end{equation}
\noindent where
\begin{equation}
\omega_D^2 = {\frac {e a_D} {p_0 \beta_e \gamma_e}}. \label{DefocusFreq}
\end{equation}
\noindent In addition to the above, the coordinate transformation at the polygon bend when passing to the new rotated coordinate system needs to be specified. The latter can be written as
\begin{equation}
X_e = X_c + {\frac {X_0 - X_c} {\cos \Theta_p - P_0 \sin \Theta_p / \beta_e \gamma_e}}, \nonumber
\end{equation}
\begin{equation}
P_e = \beta_e \gamma_e \tan {\left[ \Theta_p + \arctan {\left( {\frac {P_0} {\beta_e \gamma_e}} \right)} \right]}. \label{PolyBend}
\end{equation}

Once the reference trajectory has been found the corresponding contributions to the total Hamiltonian (\ref{SHamiltdec}) can be written as follows
\begin{equation}
\widetilde{H}_0 = - \beta \gamma + {\frac {Z} {A E_p}} {\left( {\frac {{\rm d} \Delta E} {{\rm d} s}} \right)} \int {\rm d} \Theta \sin \phi (\Theta), \label{SHamiltde0}
\end{equation}
\begin{equation}
\widetilde{H}_1 = - {\left( \beta \gamma - \beta_e \gamma_e \right)} K \widetilde{x}, \label{SHamiltde1}
\end{equation}
\begin{equation}
\widetilde{H}_2 = {\frac {1} {2 \beta \gamma}} {\left( {\widetilde{P}}_x^2 + {\widetilde{P}}_z^2 \right)} + {\frac {1} {2}} {\left[ {\left( g + \beta_e \gamma_e K^2 \right)} {\widetilde{x}}^2 - g {\widetilde{z}}^2 \right]}, \label{SHamiltde2}
\end{equation}
\begin{equation}
\widetilde{H}_3 = {\frac {K {\widetilde{x}}} {2 \beta \gamma}} {\left( {\widetilde{P}}_x^2 + {\widetilde{P}}_z^2 \right)} + {\frac {K g} {6}} {\left( 2 {\widetilde{x}}^3 - 3 \widetilde{x} {\widetilde{z}}^2 \right)}, \label{SHamiltdc3}
\end{equation}
\begin{equation}
\widetilde{H}_4 = {\frac {{\left( {\widetilde{P}}_x^2 + {\widetilde{P}}_z^2 \right)}^2} {8 \beta^3 \gamma^3}} - {\frac {K^2 g} {24}} {\widetilde{z}}^4. \label{SHamiltdc4}
\end{equation}
\noindent Here, we have introduced the following notation
\begin{equation}
g = {\frac {q a_1} {p_0}}. \label{Notatgxz}
\end{equation}
\noindent Moreover, $Z$ is the charge state of the accelerated particle, $A$ is the mass ratio with respect to the proton mass in the case of ions, and $\phi (\Theta)$ is the phase of the RF. For a lepton accelerator like EMMA, $A = Z = 1$. In addition, $({\rm d} \Delta E / {\rm d} s)$ is the energy gain per unit longitudinal distance $s$, which in thin lens approximation scales as $\Delta E / \Delta s$, where $\Delta s$ is the length of the cavity. It is convenient to pass to new scaled variables as follows
\begin{equation}
{\widetilde{p}}_u = {\frac {{\widetilde{P}}_u} {\beta_e \gamma_e}}, \qquad h = {\frac {\gamma} {\beta_e^2 \gamma_e}}, \label{Scaledvarb}
\end{equation}
\begin{equation}
\tau = \beta_e \Theta, \qquad \Gamma_e = {\frac {\beta \gamma} {\beta_e \gamma_e}} =  {\sqrt{\beta_e^2 h^2 - {\frac {1} {\beta_e^2 \gamma_e^2}}}}. \label{Scaledvar1}
\end{equation}
\noindent Thus, expressions (\ref{SHamiltde0}) -- (\ref{SHamiltdc4}) become
\begin{equation}
\widetilde{H}_0 = - \Gamma_e + {\frac {Z} {A \beta_e^2 E_e}} {\left( {\frac {{\rm d} \Delta E} {{\rm d} s}} \right)} \int {\rm d} \tau \sin \phi (\tau), \label{SaHamiltde0}
\end{equation}
\begin{equation}
\widetilde{H}_1 = - {\left( \Gamma_e - 1 \right)} K \widetilde{x}, \label{SaHamiltde1}
\end{equation}
\begin{equation}
\widetilde{H}_2 = {\frac {1} {2 \Gamma_e}} {\left( {\widetilde{p}}_x^2 + {\widetilde{p}}_z^2 \right)} + {\frac {1} {2}} {\left[ {\left( g_e + K^2 \right)} {\widetilde{x}}^2 - g_e {\widetilde{z}}^2 \right]}, \label{SaHamiltde2}
\end{equation}
\begin{equation}
\widetilde{H}_3 = {\frac {K {\widetilde{x}}} {2 \Gamma_e}} {\left( {\widetilde{p}}_x^2 + {\widetilde{p}}_z^2 \right)} + {\frac {K g_e} {6}} {\left( 2 {\widetilde{x}}^3 - 3 \widetilde{x} {\widetilde{z}}^2 \right)}, \label{SaHamiltdc3}
\end{equation}
\begin{equation}
\widetilde{H}_4 = {\frac {{\left( {\widetilde{p}}_x^2 + {\widetilde{p}}_z^2 \right)}^2} {8 \Gamma_e^3}} - {\frac {K^2 g_e} {24}} {\widetilde{z}}^4, \label{SaHamiltdc4}
\end{equation}
\begin{equation}
E_p = m_{p0} c^2, \qquad \qquad g_e = {\frac {g} {\beta_e \gamma_e}}. \label{SaHamiltdc41}
\end{equation}

The longitudinal part of the reference orbit can be isolated via a canonical transformation
\begin{equation}
F_2 {\left( \widetilde{x}, \widetilde{\widetilde{p}}_x, \widetilde{z}, \widetilde{\widetilde{p}}_z, \tau, \eta; s \right)} = \widetilde{x} \widetilde{\widetilde{p}}_x + \widetilde{z} \widetilde{\widetilde{p}}_z + {\left( \tau + s \right)} {\left( \eta + {\frac {1} {\beta_e^2}} \right)}, \label{Rogenfunc}
\end{equation}
\begin{equation}
\sigma = \tau + s, \qquad \qquad \eta = h - {\frac {1} {\beta_e^2}}, \label{Rodefinition}
\end{equation}
\noindent where $\sigma$ is the new longitudinal variable and $\eta$ is the energy deviation with respect to the energy $\gamma_e$ of the reference particle.

\section{Dispersion and Betatron Motion}

The (linear and higher order) dispersion can be introduced via a canonical transformation aimed at canceling the first order Hamiltonian ${\widetilde{H}}_1$ in all orders of $\eta$. The explicit form of the generating function is
\begin{equation}
G_2 {\left( {\widetilde{x}}, {\widehat{p}}_x, {\widetilde{z}}, {\widehat{p}}_z, \sigma, \widehat{\eta}; s \right)} = \sigma {\widehat{\eta}} + {\widetilde{z}} {\widehat{p}}_z + {\widetilde{x}} {\widehat{p}}_x \nonumber
\end{equation}
\begin{equation}
+ \sum \limits_{k=1}^{\infty} {\widehat{\eta}}^k {\left[ {\widetilde{x}} {\cal X}_k (s) -{\widehat{p}}_x {\cal P}_k (s) + {\cal S}_k (s) \right]}, \label{Dispgenfunc}
\end{equation}
\begin{equation}
{\widetilde{x}} = \widehat{x} + \sum \limits_{k=1}^{\infty} {\widehat{\eta}}^k {\cal P}_k, \qquad \qquad {\widetilde{p}}_x = {\widehat{p}}_x + \sum \limits_{k=1}^{\infty} {\widehat{\eta}}^k {\cal X}_k, \label{Canvardisp}
\end{equation}
\begin{equation}
\sigma = \widehat{\sigma} + \sum \limits_{k=1}^{\infty} k {\widehat{\eta}}^{k-1} {\left( {\cal P}_k {\widehat{p}}_x - {\cal X}_k {\widehat{x}} \right)} \nonumber
\end{equation}
\begin{equation}
- \sum \limits_{k=1}^{\infty} k {\widehat{\eta}}^{k-1} {\left( {\cal S}_k + {\cal X}_k \sum \limits_{m=1}^{\infty} {\widehat{\eta}}^m {\cal P}_m \right)}. \label{Canvarlondisp}
\end{equation}
\noindent Equating terms of the form ${\widehat{x}} {\widehat{\eta}}^n$ and ${\widehat{p}}_x {\widehat{\eta}}^n$ in the new transformed Hamiltonian, we determine order by order the conventional (first order) and higher order dispersions. The first order in ${\widehat{\eta}}$ (terms proportional to ${\widehat{x}} {\widehat{\eta}}$ and ${\widehat{p}}_x {\widehat{\eta}}$) yields the well-known result
\begin{equation}
{\cal P}_1^{\prime} = {\cal X}_1, \qquad \qquad {\cal X}_1^{\prime} + {\left( g_e + K^2 \right)} {\cal P}_1 = K. \label{Dispdiffequa}
\end{equation}
\noindent Since in the case of vanishing betatron motion ${\left( {\widehat{x}} = 0, \quad {\widehat{p}}_x = 0 \right)}$ the new longitudinal coordinate $\widehat{\sigma}$ should not depend on the new longitudinal canonical conjugate variable $\widehat{\eta}$, the second sum in equation (\ref{Canvarlondisp}) must be identically zero. We readily obtain ${\cal S}_1 = 0$, and
\begin{equation}
{\cal S}_2 = - {\frac {{\cal X}_1 {\cal P}_1} {2}}. \label{SecondordS}
\end{equation}
\noindent In second order we have
\begin{equation}
{\cal P}_2^{\prime} = {\cal X}_2 - {\cal X}_1 + K {\cal X}_1 {\cal P}_1, \label{Dispdiffequap2}
\end{equation}
\begin{equation}
{\cal X}_2^{\prime} + {\left( g_e + K^2 \right)} {\cal P}_2 = - K g_e {\cal P}_1^2 - {\frac {K {\cal X}_1^2} {2}} - {\frac {K} {2 \gamma_e^2}}, \label{Dispdiffequax2}
\end{equation}
\noindent and in addition the function ${\cal S}_3 (s)$ is expressed as
\begin{equation}
{\cal S}_3 = - {\frac {1} {3}} {\left( {\cal X}_1 {\cal P}_2 + 2 {\cal X}_2 {\cal P}_1 \right)}. \label{ThirdordS}
\end{equation}

Close inspection of equations (\ref{Dispdiffequa}), (\ref{Dispdiffequap2}) and (\ref{Dispdiffequax2}) shows that ${\cal P}_1$ is the well-known linear dispersion function, ${\cal P}_2$ stands for a second order dispersion and so on. Up to third order in ${\widehat{\eta}}$ the new Hamiltonian describing the longitudinal motion and the linear transverse motion acquires the form
\begin{equation}
\widehat{H}_0 = - {\frac {{\widetilde{\cal K}}_1 {\widehat{\eta}}^2} {2}} + {\frac {{\widetilde{\cal K}}_2 {\widehat{\eta}}^3} {3}} + {\frac {Z} {A \beta_e^2 E_e}} {\left( {\frac {{\rm d} \Delta E} {{\rm d} s}} \right)} \int {\rm d} \tau \sin \phi (\tau), \label{Hamilpsliptde0}
\end{equation}
\begin{equation}
\widehat{H}_2 = {\frac {1} {2}} {\left( {\widehat{p}}_x^2 + {\widehat{p}}_z^2 \right)} + {\frac {1} {2}} {\left[ {\left( g_e + K^2 \right)} {\widehat{x}}^2 - g_e {\widehat{z}}^2 \right]}, \label{SbetHamiltde2}
\end{equation}
\noindent where
\begin{equation}
{\widetilde{\cal K}}_1 = K {\cal P}_1 - {\frac {1} {\gamma_e^2}} \qquad {\widetilde{\cal K}}_2 = {\frac {K {\cal P}_1} {\gamma_e^2}} - K {\cal P}_2 - {\frac {{\cal X}_1^2} {2}} - {\frac {3} {2 \gamma_e^2}}. \label{Phaseslips}
\end{equation}

For the sake of generality, let us consider a Hamiltonian of the type
\begin{equation}
\widehat{H}_b = \sum \limits_{u=(x,z)} {\left[ {\frac {{\cal F}_u} {2}} {\widehat{p}}_u^2 + {\cal R}_u {\widehat{u}} {\widehat{p}}_u + {\frac {{\cal G}_u} {2}} {\widehat{u}}^2 \right]}. \label{SbetHamiltde21}
\end{equation}
\noindent A generic Hamiltonian of the type (\ref{SbetHamiltde21}) can be transformed to the normal form
\begin{equation}
{\mathcal{H}}_b = \sum \limits_{u=(x, z)} {\frac {\chi_u^{\prime}} {2}} {\left( {\overline{P}}_u^2 + {\overline{U}}^2 \right)}, \label{SbetHamiltnor}
\end{equation}
\noindent by means of a canonical transformation specified by the generating function
\begin{equation}
{\mathcal{F}}_2 {\left( \widehat{x}, {\overline{P}}_x, \widehat{z}, {\overline{P}}_z; s \right)} = \sum \limits_{u=(x, z)} {\left( {\frac {\widehat{u} {\overline{P}}_u} {\sqrt{\beta_u}}} - {\frac {\alpha_u {\widehat{u}}^2} {2 \beta_u}} \right)}. \label{Genfuncnor}
\end{equation}
\noindent Here the prime implies differentiation with respect to the longitudinal variable $s$. The old and the new canonical variables are related through the expressions
\begin{equation}
\widehat{u} = {\overline{U}} \sqrt{\beta_u}, \qquad \qquad {\widehat{p}}_u = {\frac {1} {\sqrt{\beta_u}}} {\left( {\overline{P}}_u - \alpha_u {\overline{U}} \right)}. \label{Newoldnor}
\end{equation}
\noindent The phase advance $\chi_u (s)$ and the generalized Twiss parameters $\alpha_u (s)$, $\beta_u (s)$ and $\gamma_u (s)$ are defined as
\begin{equation}
\chi_u^{\prime} = {\frac {{\rm d} \chi_u} {{\rm d} s}} = {\frac {{\cal F}_u} {\beta_u}}, \label{Phaseadv}
\end{equation}
\begin{equation}
\alpha_u^{\prime} = {\frac {{\rm d} \alpha_u} {{\rm d} s}} = {\cal G}_u \beta_u - {\cal F}_u \gamma_u, \label{AlphaTwiss}
\end{equation}
\begin{equation}
\beta_u^{\prime} = {\frac {{\rm d} \beta_u} {{\rm d} s}} = - 2 {\cal F}_u \alpha_u + 2 {\cal R}_u \beta_u. \label{BetaTwiss}
\end{equation}
\noindent The third Twiss parameter $\gamma_u (s)$ is introduced via the well-known expression
\begin{equation}
\beta_u \gamma_u - \alpha_u^2 = 1. \label{GammaTwiss}
\end{equation}
\noindent The corresponding betatron tunes are determined according to the expression
\begin{equation}
\nu_u = {\frac {N_p} {2 \pi}} \int \limits_{s}^{s + L_p} {\frac {{\rm d} \theta {\cal F}_u (\theta)} {\beta_u (\theta)}}. \label{Bettunes}
\end{equation}

\begin{figure}
\begin{center}
\resizebox{80mm}{!}
{\includegraphics{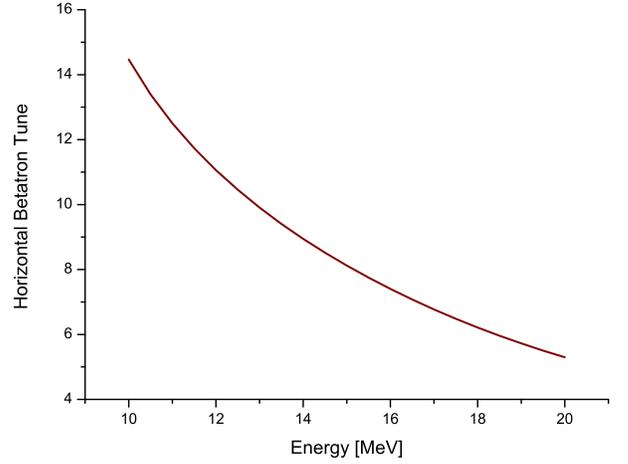}}
\caption{\label{fig:hortune} Horizontal betatron tune for the EMMA ring as a function of energy.}
\end{center}
\end{figure}

\begin{figure}
\begin{center}
\resizebox{80mm}{!}
{\includegraphics{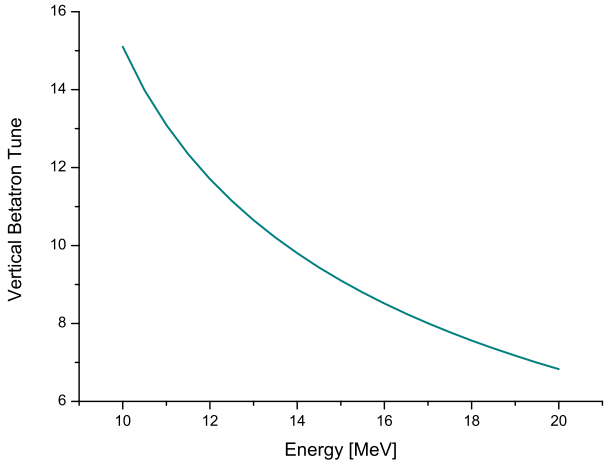}}
\caption{\label{fig:vertune} Vertical betatron tune for the EMMA ring as a function of energy.}
\end{center}
\end{figure}

Typical dependence of the horizontal and vertical betatron tunes on energy in the EMMA non scaling FFAG is shown in Figures \ref{fig:hortune} and \ref{fig:vertune}.

\section{Acceleration in a Non Scaling FFAG Accelerator}

The process of acceleration in a non scaling FFAG accelerator can be studied by solving Hamilton's equations of motion for the longitudinal degree of freedom. The latter are obtained from the Hamiltonian (\ref{EquivHam}) supplemented by an additional term [similar to that in equation (\ref{SHamiltde0})], which takes into account the electric field of
the RF cavities. They read as
\begin{equation}
{\frac {{\rm d} \Theta} {{\rm d} s}} = - {\frac {\gamma} {\sqrt{\beta^2 \gamma^2 - P^2}}}, \label{Rodiffequat}
\end{equation}
\begin{equation}
{\frac {{\rm d} \gamma} {{\rm d} s}} = - {\frac {Z e U_c} {2 A E_p}} \sum \limits_{k=1}^{N_c} \delta_p {\left(s - s_k \right)} \sin {\left( {\frac {\omega_c \Theta} {c}} - \varphi_k \right)}. \label{Rodiffequag}
\end{equation}
\noindent Here $U_c$ is the cavity voltage, $\omega_c$ is the RF frequency, $N_c$ is the number of cavities and $\varphi_k$ is the corresponding cavity phase.

\begin{figure}
\begin{center}
\resizebox{80mm}{!}
{\includegraphics{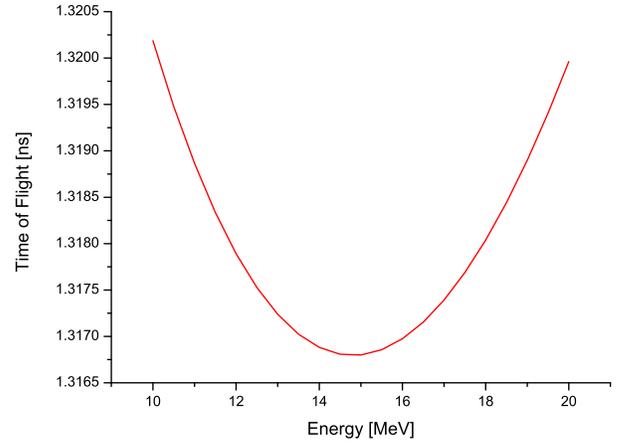}}
\caption{\label{fig:tof} Time of flight as a function of energy for a single 0.394481 meter EMMA cell.}
\end{center}
\end{figure}

One could use the results obtained in the previous section with the additional requirement that the phase slip coefficient ${\widetilde{\cal K}}_1$ averaged over one period vanishes. Instead, we shall use an equivalent but more illustrative approach. The path length in a FFAG arc and therefore the time of flight $\Theta$ is often well approximated as a quadratic function of energy. The acceleration process is then described by a longitudinal Hamiltonian, which contains terms proportional to the zero-order (conventional phase slip) factor and first-order phase slip factor. It usually suffices to take into account only terms to second order in the energy deviation
\begin{equation}
\Theta = \Theta_0 + 2 {\cal A} \gamma_m \gamma - {\cal A} \gamma^2, \label{TOFEnerg}
\end{equation}
\noindent as suggested by Figure \ref{fig:tof}.

Here $\gamma_m$ corresponds to the reference energy with a minimum time of flight. Provided the time of flight $\Theta_i$ at injection energy $\gamma_i$ and the time of flight $\Theta_m$ at reference energy $\gamma_m$ are known, the constants entering equation (\ref{TOFEnerg}) can be expressed as
\begin{equation}
{\cal A} = {\frac {\Theta_m - \Theta_i} {{\left( \gamma_m - \gamma_i \right)}^2}}, \qquad \qquad \Theta_0 = \Theta_m - {\cal A} \gamma_m^2. \label{ConstTOF}
\end{equation}
\noindent Next, we pass to a new variable
\begin{equation}
{\widehat{\gamma}} = \gamma - \gamma_m, \qquad \qquad \Theta = \Theta_m - {\cal A} {\widehat{\gamma}}^2, \label{NewVarTOF}
\end{equation}
\noindent similar to the variable ${\widehat{\eta}}$ introduced in the previous section. Then, Hamilton's equation of motion (\ref{Rodiffequat}) can be rewritten in an equivalent form
\begin{equation}
{\frac {{\rm d} \Theta} {{\rm d} s}} = {\frac {\Theta_m} {L_p}} - {\frac {{\cal A} {\widehat{\gamma}}^2} {L_p}}, \label{TOFequat}
\end{equation}

In what follows, it is convenient to introduce a new phase ${\widetilde{\varphi}}$ and the azimuthal angle $\theta$ along the machine circumference as an independent variable according to the relations
\begin{equation}
{\rm d} s = R {\rm d} \theta, \qquad {\widetilde{\varphi}} = {\frac {\omega_c \Theta} {c}}, \qquad R = {\frac {N_L L_p} {2 \pi}}. \label{TOFrelat}
\end{equation}
\noindent It is straightforward to verify (see the averaging procedure below) that the necessary condition to have acceleration is
\begin{equation}
{\frac {\omega_c N_L {\left| \Theta_m \right|}} {2 \pi c}} = h, \label{Accelcond}
\end{equation}
\noindent where $h$ is an integer (a harmonic number). Averaging Hamilton's equations of motion
\begin{equation}
{\frac {{\rm d} {\widetilde{\varphi}}} {{\rm d} \theta}} = - h - h a {\widehat{\gamma}}^2, \qquad \qquad a = {\frac {\cal A} {\left| \Theta_m \right|}}, \label{AccelHamiph}
\end{equation}
\begin{equation}
{\frac {{\rm d} {\widehat{\gamma}}} {{\rm d} \theta}} = - {\frac {Z e U_c} {2 A E_p}} \sum \limits_{k=1}^{N_c} \delta_p {\left( \theta - \theta_k \right)} \sin {\left( {\widetilde{\varphi}} - \varphi_k \right)}, \label{AccelHamiga}
\end{equation}
\noindent we rewrite them in a simpler form as
\begin{equation}
{\frac {{\rm d} \varphi} {{\rm d} \theta}} = h a {\widehat{\gamma}}^2, \qquad \qquad {\frac {{\rm d} {\widehat{\gamma}}} {{\rm d} \theta}} = \lambda \sin \varphi, \label{AccelHamipha}
\end{equation}
\noindent where
\begin{equation}
\varphi = - {\widetilde{\varphi}} - h \theta + \psi_0, \qquad \qquad \lambda = {\frac {Z e U_c {\cal D}} {4 \pi A E_p}}, \label{Accelcowffic}
\end{equation}
\begin{equation}
{\cal D} = {\sqrt{{\cal A}_c^2 + {\cal A}_s^2}}, \qquad \qquad \psi_0 = \arctan {\left( {\frac {{\cal A}_s} {{\cal A}_c}} \right)}, \label{Accelcowffic1}
\end{equation}
\begin{equation}
{\cal A}_c = \sum \limits_{k=1}^{N_c} \cos {\left( h \theta_k + \varphi_k \right)}, \qquad {\cal A}_s = \sum \limits_{k=1}^{N_c} \sin {\left( h \theta_k + \varphi_k \right)}. \label{Accelcowffic2}
\end{equation}
\noindent The effective longitudinal Hamiltonian, which governs the equations of motion (\ref{AccelHamipha}) can be written as
\begin{equation}
H_0 = {\frac {h a} {3}} {\widehat{\gamma}}^3 + \lambda \cos \varphi. \label{AccelLongHam}
\end{equation}
\noindent Since the Hamiltonian (\ref{AccelLongHam}) is a constant of motion, the second Hamilton equation (\ref{AccelHamipha}) can be written as
\begin{equation}
{\frac {{\rm d} {\widehat{\gamma}}} {{\rm d} \theta}} = \pm \lambda {\sqrt{1 - {\frac {1} {\lambda^2}} {\left( H_0 - {\frac {h a} {3}} {\widehat{\gamma}}^3 \right)}^2}}. \label{AccelSecHami}
\end{equation}

\begin{figure}
\begin{center}
\resizebox{80mm}{!}
{\includegraphics{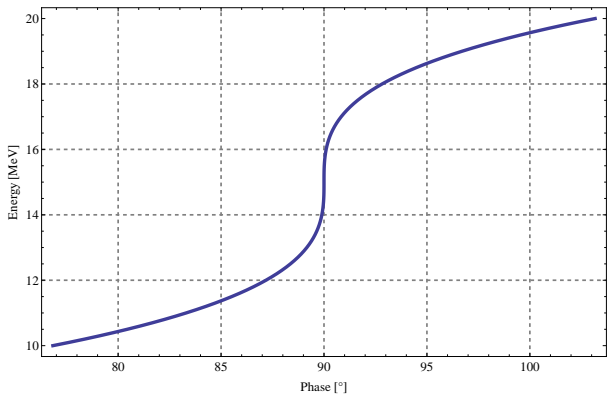}}
\caption{\label{Separatrix1} An example of the so-called serpentine acceleration for the EMMA ring for the central trajectory, where the longitudinal $H_0 = 0$. The harmonic number is assumed to be 11, with the RF wavelength 0.405m. The parameter $a$ from Eq. (\ref{AccelHamiph}) is taken to be $2.6863 10^{-5}$.}
\end{center}
\end{figure}

Let us first consider the case of the central trajectory, where $H_0 = 0$. It is of utmost importance for the so called gutter acceleration. Equation (\ref{AccelSecHami}) can be solved in a straightforward manner to give
\begin{equation}
\theta = {\frac {J} {b}} \, {}_2F_{1} {\left( {\frac {1} {6}}, {\frac {1} {2}}; {\frac {7} {6}}; J^6 \right)} - {\frac {\cal C} {b}}, \label{AccelSolCent}
\end{equation}
\noindent where
\begin{equation}
J = {\widehat{\gamma}} \, {\sqrt[3]{\frac {h a} {3 \lambda}}}, \qquad \qquad b = \lambda \, {\sqrt[3]{\frac {h a} {3 \lambda}}}, \label{AccelSolCoeff}
\end{equation}
\begin{equation}
{\cal C} = {}_2F_{1} {\left( {\frac {1} {6}}, {\frac {1} {2}}; {\frac {7} {6}}; J_i^6 \right)} J_i. \label{AccelSolCoCal}
\end{equation}
\noindent In the above expressions ${}_2F_{1} {\left( \alpha, \beta; \gamma; x \right)}$ denotes the Gauss hypergeometric function of the argument $x$. This case is illustrated in Figure \ref{Separatrix1}.

\begin{figure}
\begin{center}
\resizebox{80mm}{!}
{\includegraphics{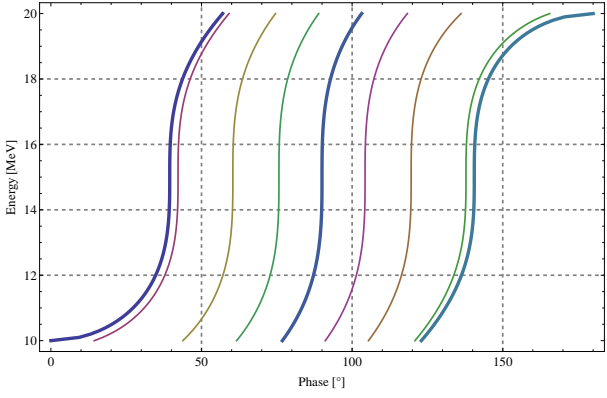}}
\caption{\label{Separatrix2} Examples of serpentine acceleration for the EMMA ring, with varying value of the longitudinal Hamiltonian. The limits of stability are given at values of the longitudinal Hamiltonian of $\pm 0.31272$, corresponding to either a 0 phase at 10MeV, or a $\pi$ phase at 20MeV.}
\end{center}
\end{figure}

In the general case where $H_0 \neq 0$, we have
\begin{equation}
\theta = {\frac {J} {b {\sqrt{a_1 c}}}} \, F_{1} {\left( {\frac {1} {3}}; {\frac {1} {2}}, {\frac {1} {2}}; {\frac {4} {3}}; {\frac {J^3} {a_1}}, - {\frac {J^3} {c}} \right)} - {\frac {{\cal C}_1} {b}}, \label{AccelSolGutt}
\end{equation}
\noindent where
\begin{equation}
a_1 = 1 + {\frac {H_0} {\lambda}}, \qquad \qquad c = 1 - {\frac {H_0} {\lambda}}, \label{AccelSolCoefg}
\end{equation}
\begin{equation}
{\cal C}_1 = {\frac {J_i} {\sqrt{a_1 c}}} \, F_{1} {\left( {\frac {1} {3}}; {\frac {1} {2}}, {\frac {1} {2}}; {\frac {4} {3}}; {\frac {J_i^3} {a_1}}, - {\frac {J_i^3} {c}} \right)}. \label{AccelSolCoCalg}
\end{equation}
\noindent Here now, $F_{1} {\left( \alpha; \beta, \gamma; \delta; x, y \right)}$ denotes the Appell hypergeometric function of the arguments $x$ and $y$. The phase portrait corresponding to the general case for a variety of values of the longitudinal Hamiltonian $H_0$ is illustrated in Figure \ref{Separatrix2}.

\section{Concluding Remarks}

Based on the Hamiltonian formalism, the synchro-betatron approach for the description of the dynamics of particles in non scaling FFAG machines has been developed. Its starting point is the specification of the static reference (closed) orbit for a fixed energy as a solution of the equations of motion in the machine reference frame. The problem of dynamical stability and acceleration is sequentially studied in the natural coordinate system associated with the reference orbit thus determined.

It has been further shown that the dependence of the path length on the energy deviation can be described in terms of higher order (nonlinear) dispersion functions. The method provides a systematic tool to determine the dispersion functions to every desired order, and represents a natural definition through constitutive equations for the resulting Twiss parameters.

The formulation thus developed has been applied to the electron FFAG machine EMMA. The transverse and longitudinal dynamics are explored and an initial attempt is made at understanding the limits of longitudinal stability of such a machine.

Unlike the conventional synchronous acceleration, the acceleration process in FFAG accelerators is an asynchronous one in which the reference particle performs nonlinear oscillations around the crest of the RF waveform. To the best of our knowledge, it is the first time that such a fully analytic theory describing the acceleration in non scaling FFAGs has been developed.

\appendix

\section{Calculation of the Reference Orbit}

The explicit solutions of the linearized equations of motion (\ref{HamEquatLin}) can be used to calculate approximately the reference orbit. To do so, we introduce a state vector
\begin{equation}
{\bf Z}_e = {\left(
\begin{array}{cc}
X_e \\
\\
P_e
\end{array} \right)}. \label{StateVec}
\end{equation}
\noindent The effect of each lattice element can be represented in a simple form as
\begin{equation}
{\bf Z}_{out} = {\widehat{\cal M}}_{el} {\bf Z}_{in} + {\bf A}_{el}. \label{ElemEffect}
\end{equation}
\noindent Here ${\bf Z}_{in}$ is the initial value of the state vector, while ${\bf Z}_{out}$ is its final value at the exit of the corresponding element. The transfer matrix ${\widehat{\cal M}}_{el}$ and the shift vector ${\bf A}_{el}$ for various lattice elements are given as follows:

1. {\it Polygon Bend}.

Within the approximation (\ref{Approx}) considered here we can linearize the second of equations (\ref{PolyBend}) and write
\begin{equation}
{\widehat{\cal M}}_p = {\left(
\begin{array}{cc}
1 / \cos \Theta_p \ \ - X_c \tan \Theta_p / {\left( \beta_e \gamma_e \cos \Theta_p \right)} \\
\\
0 \ \ \ \ \ \ \ \ \ \ 1 / \cos^2 \Theta_p
\end{array} \right)}, \nonumber
\end{equation}
\begin{equation}
{\bf A}_p  = {\left(
\begin{array}{cc}
X_c {\left(1 - 1 / \cos \Theta_p \right)} \\
\\
\beta_e \gamma_e \tan \Theta_p
\end{array} \right)}. \label{TransMatPoly}
\end{equation}

2. {\it Drift Space}.

\begin{equation}
{\widehat{\cal M}}_O = {\left(
\begin{array}{cc}
1 \ L_O / \beta_e \gamma_e \\
\\
0 \ \ \ \ \ \ \ \ \ 1
\end{array} \right)}, \qquad \qquad
{\bf A}_O  = 0, \label{TransMatDrift}
\end{equation}
\noindent where $L_O$ is the length of the drift. Every cell of the EMMA lattice includes a short drift of length $L_0$ and a long one of length $L_1$.

3. {\it Focusing Quadrupole}.

The transfer matrix can be written in a straightforward manner as
\begin{equation}
{\widehat{\cal M}}_F = {\left(
\begin{array}{cc}
\cos {\left( \omega_F L_F \right)} \ \ \ \sin {\left( \omega_F L_F \right)} / {\left( \beta_e \gamma_e \omega_F \right)} \\
\\
- \beta_e \gamma_e \omega_F \sin {\left( \omega_F L_F \right)} \ \ \ \cos {\left( \omega_F L_F \right)}
\end{array} \right)}, \label{TransMatFocus}
\end{equation}
\begin{equation}
{\bf A}_F = {\left(
\begin{array}{cc}
{\left( X_c + d_F \right)} {\left[ 1 - \cos {\left( \omega_F L_F \right)} \right]} \\
\\
\beta_e \gamma_e \omega_F {\left( X_c + d_F \right)} \sin {\left( \omega_F L_F \right)}
\end{array} \right)}, \label{TrMatrixZF}
\end{equation}
\noindent where $L_F$ is the length of the focusing quadrupole.

4. {\it Defocusing Quadrupole}.

The transfer matrix in this case can be written in analogy to the above one as
\begin{equation}
{\widehat{\cal M}}_D = {\left(
\begin{array}{cc}
\cosh {\left( \omega_D L_D \right)} \ \ \ \sinh {\left( \omega_D L_D \right)} / {\left( \beta_e \gamma_e \omega_D \right)} \\
\\
\beta_e \gamma_e \omega_D \sinh {\left( \omega_D L_D \right)} \ \ \ \cosh {\left( \omega_D L_D \right)}
\end{array} \right)}, \label{TransMatDefo}
\end{equation}
\begin{equation}
{\bf A}_D = {\left(
\begin{array}{cc}
{\left( X_c + d_D \right)} {\left[ 1 - \cosh {\left( \omega_D L_D \right)} \right]} \\
\\
- \beta_e \gamma_e \omega_D {\left( X_c + d_D \right)} \sinh {\left( \omega_D L_D \right)}
\end{array} \right)}, \label{TraMatrixZF}
\end{equation}
\noindent where $L_D$ is the length of the defocusing quadrupole.

Since the reference orbit must be a periodic function of $s$ with period $L_p$, it clearly satisfies the condition
\begin{equation}
{\bf Z}_{out} = {\bf Z}_{in} = {\bf Z}_e. \label{PeriodCond}
\end{equation}
\noindent Thus, the equation for determining the reference orbit becomes
\begin{equation}
{\bf Z}_e = {\widehat{\cal M}} {\bf Z}_e + {\bf A}, \qquad {\rm or} \qquad {\bf Z}_e = {\left( 1 - {\widehat{\cal M}} \right)}^{-1} {\bf A}. \label{RefOrbitEq}
\end{equation}
\noindent Here ${\widehat{\cal M}}$ and ${\bf A}$ are the transfer matrix and the shift vector for one period, respectively. The inverse of the matrix $1 - {\widehat{\cal M}}$ can be expressed as
\begin{equation}
{\left( 1 - {\widehat{\cal M}} \right)}^{-1} = {\frac {\cos^3 \Theta_p} {1 + {\left( 1 - {\rm Sp} {\widehat{\cal M}} \right)} \cos^3 \Theta_p}} \nonumber
\end{equation}
\begin{equation}
\times {\left(
\begin{array}{cc}
1 - {\cal M}_{22} \ \ \ {\cal M}_{12} \\
\\
{\cal M}_{21} \ \ \ 1 - {\cal M}_{11}
\end{array} \right)}. \label{TransMatInv}
\end{equation}

For the EMMA lattice in particular, the components of the one period transfer matrix and shift vector can be written explicitly as
\begin{widetext}
\begin{equation}
{\cal M}_{11} = {\frac {1} {c_p}} {\left[ c_F c_D + {\left( {\frac {\omega_D} {\omega_F}} - L_0 L_1 \omega_F \omega_D \right)} s_F s_D + {\left( L_0 + L_1 \right)} \omega_D c_F s_D - L_1 \omega_F s_F c_D \right]}, \label{TransMat11}
\end{equation}
\begin{eqnarray}
{\cal M}_{12} = {\frac {1} {\beta_e \gamma_e c_p}} {\left\{ {\left( {\frac {L_0 + L_1} {c_p}} - X_c t_p \right)} c_F c_D + {\left[ {\left( L_0 L_1 \omega_F \omega_D - {\frac {\omega_D} {\omega_F}} \right)} X_c t_p - {\frac {\omega_F L_1} {\omega_D c_p}} \right]} s_F s_D \right.} \nonumber
\end{eqnarray}
\begin{equation}
{\left. + {\left[ {\frac {1} {\omega_D c_p}} - {\left( L_0 + L_1 \right)} \omega_D X_c t_p \right]} c_F s_D + {\left( {\frac {1} {\omega_F c_p}} + L_1 \omega_F X_c t_p - {\frac {L_0 L_1 \omega_F} {c_p}} \right)} s_F c_D \right\}}, \label{TransMat12}
\end{equation}
\begin{equation}
{\cal M}_{21} = - {\frac {\beta_e \gamma_e} {c_p}} {\left( \omega_F s_F c_D + L_0 \omega_F \omega_D s_F s_D - \omega_D c_F s_D \right)}, \label{TransMat21}
\end{equation}
\begin{equation}
{\cal M}_{22} = {\frac {1} {c_p}} {\left[ {\frac {c_F c_D} {c_p}} + {\left( L_0 \omega_F \omega_D X_c t_p - {\frac {\omega_F} {\omega_D c_p}} \right)} s_F s_D + \omega_F {\left( X_c t_p - {\frac {L_0} {c_p}} \right)} s_F c_D - \omega_D X_c t_p c_F s_D \right]}, \label{TransMat22}
\end{equation}
\begin{eqnarray}
A_1 = X_c + d_F + {\left( d_D - d_F \right)} {\left( c_F - L_1 \omega_F s_F \right)} + {\left( {\frac {X_c} {c_p}} + d_D \right)} \nonumber
\end{eqnarray}
\begin{eqnarray}
\times {\left[ L_1 \omega_F s_F c_D - c_F c_D - {\left( L_0 + L_1 \right)} \omega_D c_F s_D - {\frac {\omega_D s_F s_D} {\omega_F}} + L_0 L_1 \omega_F \omega_D s_F s_D \right]} \nonumber
\end{eqnarray}
\begin{equation}
+ t_p {\left[ {\left( L_0 + L_1 \right)} c_F c_D + {\frac {c_F s_D} {\omega_D}} + {\frac {s_F c_D} {\omega_F}} - {\frac {L_1 \omega_F s_F s_D} {\omega_D}} - L_0 L_1 \omega_F s_F c_D \right]}, \label{ShiftVec1}
\end{equation}
\begin{eqnarray}
A_2 = - \beta_e \gamma_e \omega_F {\left( d_D - d_F \right)} s_F + \beta_e \gamma_e {\left( {\frac {X_c} {c_p}} + d_D \right)} {\left( \omega_F s_F c_D + \omega_F \omega_D L_0 s_F s_D - \omega_D c_F s_D \right)} \nonumber
\end{eqnarray}
\begin{equation}
+ \beta_e \gamma_e t_p {\left( c_F c_D - {\frac {\omega_F s_F s_D} {\omega_D}} - \omega_F L_0 s_F c_D \right)}. \label{ShiftVec2}
\end{equation}
\end{widetext}
\noindent For the sake of brevity, the following notations
\begin{equation}
c_p = \cos \Theta_p, \quad c_F = \cos {\left( \omega_F L_F \right)}, \quad c_D = \cosh {\left( \omega_D L_D \right)}, \label{NotationAp1}
\end{equation}
\begin{equation}
t_p = \tan \Theta_p, \quad s_F = \sin {\left( \omega_F L_F \right)}, \quad s_D = \sinh {\left( \omega_D L_D \right)}, \label{NotationAp2}
\end{equation}
\noindent have been introduced in the final expressions for the components of the one period transfer matrix and shift vector.

\section{Phase Stability in FFAGs}

To study the stability of the serpentine acceleration in FFAG accelerators, we write the longitudinal Hamiltonian (\ref{AccelLongHam}) in an equivalent form
\begin{equation}
H_0 = \lambda {\left( J^3 + \cos \varphi \right)}. \label{ApendLongHam}
\end{equation}
\noindent Hamilton's equations of motion can be written as
\begin{equation}
{\frac {{\rm d} \varphi} {{\rm d} \theta}} = 3 b J^2, \qquad \qquad {\frac {{\rm d} J} {{\rm d} \theta}} = b \sin \varphi. \label{ApendHamEqu}
\end{equation}
\noindent Let $\varphi_a (\theta)$ and $J_a (\theta)$ be the exact solution of equations (\ref{ApendHamEqu}) described already in Section V. Let us further denote by $\varphi_1$ and $J_1$ a small deviation about this solution such that $\varphi = \varphi_a + \varphi_1$ and $J = J_a + J_1$. Then, the linearized equations of motion governing the evolution of $\varphi_1$ and $J_1$ are
\begin{equation}
{\frac {{\rm d} \varphi_1} {{\rm d} \theta}} = 6 b J_a J_1, \qquad \qquad {\frac {{\rm d} J_1} {{\rm d} \theta}} = b \varphi_1 \cos \varphi_a. \label{ApendHamEqulin}
\end{equation}
\noindent The latter should be solved provided the constraint
\begin{equation}
3 J_a^2 J_1 - \varphi_1 \sin \varphi_a = 0, \label{ApendLongConstr}
\end{equation}
\noindent following from the Hamiltonian (\ref{ApendLongHam}) holds. Differentiating the second of equations (\ref{ApendHamEqulin}) with respect to $\theta$ and eliminating $\varphi_1$, we obtain
\begin{equation}
{\frac {{\rm d}^2 J_1} {{\rm d} \theta^2}} - {\frac {6 b^2 H_0} {\lambda}} J_a J_1 + 15 b^2 J_a^4 J_1 = 0. \label{ApendHamEqulinJ}
\end{equation}
\noindent Next, we examine the case of separatrix acceleration with $H_0 = 0$. In Section V we showed that to a good accuracy the energy gain ${\left[ J_a (\theta) = b \theta + J_i \right]}$ is linear in the azimuthal variable $\theta$. Therefore, equation (\ref{ApendHamEqulinJ}) can be written as
\begin{equation}
{\frac {{\rm d}^2 J_1} {{\rm d} J_a^2}} + 15 J_a^4 J_1 = 0. \label{ApendHamEquJH0}
\end{equation}

\begin{figure}
\begin{center}
\resizebox{80mm}{!}
{\includegraphics{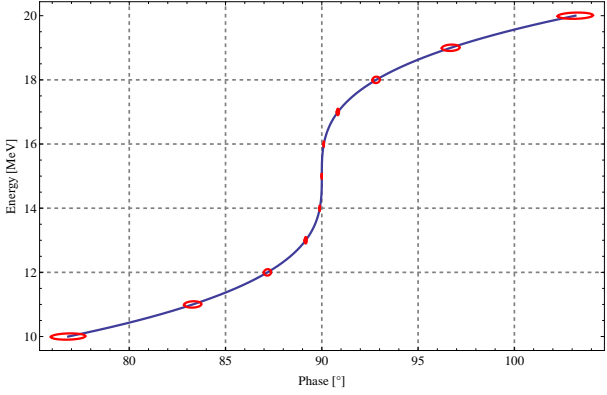}}
\caption{\label{fig:epsart} Phase stability of the standard EMMA ring, for the central trajectory at $H_0 = 0$. The errors are given as 0.1MeV in energy and $1.3^{\rm o}$ in phase.}
\end{center}
\end{figure}

\noindent The latter possesses a simple solution of the form
\begin{equation}
J_1 = {\sqrt{\left| J_a \right|}} {\left[ C_1 {\cal J}_{1/6} {\left( {\sqrt{\frac {5} {3}}} {\left| J_a \right|}^3 \right)} + C_2 {\cal Y}_{1/6} {\left( {\sqrt{\frac {5} {3}}} {\left| J_a \right|}^3 \right)} \right]}, \label{ApendHamEquSolJ}
\end{equation}
\noindent where ${\cal J}_{\alpha} (z)$ and ${\cal Y}_{\alpha} (z)$ stand for the Bessel functions of the first and second kind, respectively. In addition the constants $C_1$ and $C_2$ should be determined taking into account the initial conditions
\begin{equation}
{\frac {{\rm d} J_1 {\left( J_i \right)} } {{\rm d} J_a}} = \varphi_1 {\left( J_i \right)} \cos \varphi_i, \qquad J_1 {\left( J_i \right)} = J_{1i}. \label{ApendSolJInCond}
\end{equation}
\noindent

\newpage 
\bibliography{apssamp}

\end{document}